\documentstyle[12pt,epsf]{article}
\newcommand{\bea}{\begin{eqnarray}}
\newcommand{\eea}{\end{eqnarray}}
\newcommand{\be}{\begin{equation}}
\newcommand{\ee}{\end{equation}}

\begin{document}
\begin{Large}
\begin{center}
{\bf A STRONG ELECTROWEAK SECTOR AT FUTURE LINEAR
COLLIDERS:\\
 COMPARISON OF MODELS
}
\end{center}
\end{Large}
\vskip 1.0cm
\centerline {R. Casalbuoni$\;^{a,b)}$,
A. Deandrea$\;^{c)}$,
S. De Curtis$\;^{b)}$}
\centerline {D. Dominici$\;^{a,b)}$, R. Gatto$\;^{d)}$}
\vskip 0.5cm

\noindent
$a)$ {\it Dipartimento di Fisica, Univ. di Firenze, I-50125 Firenze, Italia.}
\hfill\break\noindent
$b)$ {\it I.N.F.N., Sezione di Firenze, I-50125 Firenze, Italia.}
\hfill\break\noindent
$c)$ {\it Centre de Phys. Th\'eorique, CNRS, Luminy, Case
907,
F-13288
Marseille
Cedex 9, France.}
\hfill\break\noindent
$d)$ {\it D\'ept. de Phys. Th\'eorique, Univ. de Gen\`eve, CH-1211 Gen\`eve
4, Suisse.}
\vskip 1cm

\def\lmu{{{{\bf L}}_\mu}}
\def \LL{{{\bf L}_\mu}}
\def\rmu{{{ \bf R}_\mu}}
\def\rmus{{\cal R}^\mu}
\def\gs{{g''}}
\def\lq{\left [}
\def\rq{\right ]}
\def\dmu{{\partial_\mu}}
\def\dnu{{\partial_\nu}}
\def\dmus{{\partial^\mu}}
\def\dnus{{\partial^\nu}}
\def\gp{g'}
\def\gpt{{{g}^\prime}}
\def\gptd{ g^{\prime 2}}
\def\eps{{\epsilon}}
\def\tr{{ {tr}}}
\def\V{{\cal{V}}}
\def\W{{\bf{W}}}
\def\Wt{\tilde{{W}}}
\def\Y{{\bf{Y}}}
\def\Yt{\tilde{{Y}}}
\def\tW{\tilde W}
\def\tY{\tilde Y}
\def\tL{\tilde L}
\def\tR{\tilde R}
\def\L{{\cal L}}
\def\s{s_\theta}
\def\c{c_\theta}
\def\gt{\tilde g}
\def\et{\tilde e}
\def\At{\tilde A}
\def\Zt{\tilde Z}
\def\Wpt{\tilde W^+}
\def\Wmt{\tilde W^-}
\def\de{\partial}
\def\eps{\epsilon}
\def\nn{\nonumber}
\def\dd{\displaystyle}
\def\ct{c_\theta}
\def\st{s_\theta}
\def\cdt{c_{2\theta}}
\def\sdt{s_{2\theta}}
\def\qq{<{\overline u}u>}
\def\Lt{{{\tilde L}}}
\def\Rt{{{\tilde R}}}
\def\st{\tilde s_\theta}
\def\ct{\tilde c_\theta}
\def\gt{\tilde g}
\def\et{\tilde e}
\def\A{\bf A}
\def\Z{\bf Z}
\def\Wpt{\tilde W^+}
\def\Wmt{\tilde W^-}
\def\nn{\nonumber}

\section{Introduction}
In this work we have studied the phenomenological properties at future
$e^+e^-$ linear colliders of two models of strong electroweak symmetry
breaking, characterized by the presence of new spin one strong interacting
resonances.

The first
model we have considered is the so called BESS model \cite{bess}.
This is an effective lagrangian
parameterization of the symmetry breaking mechanism,
based on a symmetry $G=SU(2)_L\times SU(2)_R$ broken down to
$SU(2)_{L+R}$.
New vector
particles are introduced as gauge bosons associated to a hidden $H'=SU(2)_V$.
The symmetry group of the theory becomes $G'=G\otimes H'$. It breaks down
spontaneously to $H_D=SU(2)$, which is the diagonal subgroup of $G'$. This
gives rise to six Goldstone bosons. Three are absorbed by the new vector
particles while the other three give mass to the SM gauge bosons, after the
gauging of the subgroup $SU(2)_L\otimes U(1)_Y\subset G$.
The general procedure
for building models with vector and axial vector
resonances  is discussed in \cite{assiali}.

The parameters of the BESS model
are  the mass of
these new bosons $M_V$, their self coupling
$\gs$,  and a third parameter $b$ whose strength characterizes
the direct couplings of the new vectors $V$ to the
fermions. However due to the mixing of the $V$ bosons
with $W$ and $Z$, the new particles are coupled to the
fermions also when $b=0$. The parameter $\gs$ is expected
to be large due to the fact that these new gauge bosons
are thought of as bound states from a strongly interacting electroweak
sector. By taking the formal $b\rightarrow 0$ and
$\gs\rightarrow\infty$ limits, the new bosons decouple
and the standard model (SM) is recovered. By considering only
the limit $M_V\rightarrow\infty$ they do not decouple.

The already existing data (LEP, CDF, SLC, etc.) allow us to get bounds on the
parameter space $(b, g/\gs)$, as shown in Fig. 1. The bounds do not depend on the
mass $M_V$ at least for values above $\approx 300~GeV$.
\vskip.5cm
\begin{figure}
\epsfysize=10truecm
\centerline{\epsffile{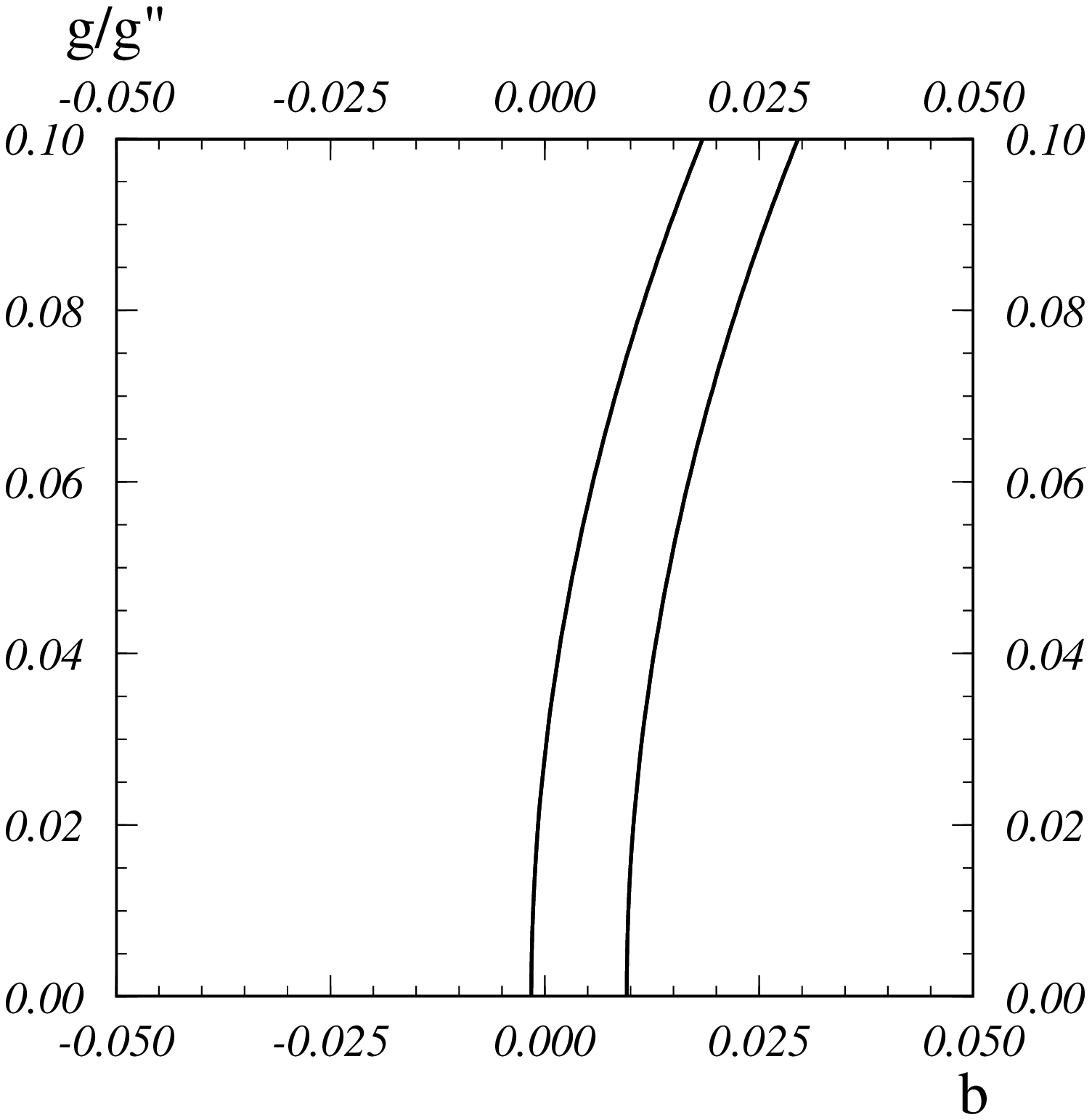}}
\smallskip
\noindent
{\bf Fig. 1} - {\it BESS model 90\% C.L. contour on the plane ($b$, $g/g''$)
obtained by
comparing the theoretical predictions
to the experimental data. The allowed region is between the
curves.}
\label{bessbo}
\end{figure}


The second model is an extension of the previous one. It is called
degenerate BESS model \cite{dege}. We start again from the global
symmetry group of the theory $G=SU(2)_L \otimes SU(2)_R$, spontaneously
broken to $SU(2)_{L+R}$. The new vector and axial vector
bosons correspond to the gauge bosons associated to a hidden symmetry,
$H'=SU(2)_L \otimes SU(2)_R$. The symmetry group of the theory becomes
$G'=G\otimes H'$. It breaks down spontaneously to $H_D$ (the same as in
 BESS) and this gives rise to
nine Goldstones. Six of these are absorbed by the vector and
axial vector bosons, and the other three  give masses to the
SM gauge bosons.
The model includes two new triplets of vector particles ($L^\pm$, $L_3$,
$R^\pm$,
$R_3$). The parameters of
the model are a new gauge coupling constant $g''$ and a mass parameter $M$,
which is the common mass of all the new vector and axial-vector particles,
when we neglect the electroweak corrections.
Contrarily to the BESS model, in this case we have decoupling in the limit
$M\to\infty$.
In the charged sector the fields $R^\pm$ are unmixed for any value of
${g''}$, and therefore they are not coupled to fermions.

The actual data give the bounds in the plane $(M, g/\gs)$ illustrated in Fig.
2.

\begin{figure}
\epsfysize=10truecm
\centerline{\epsffile{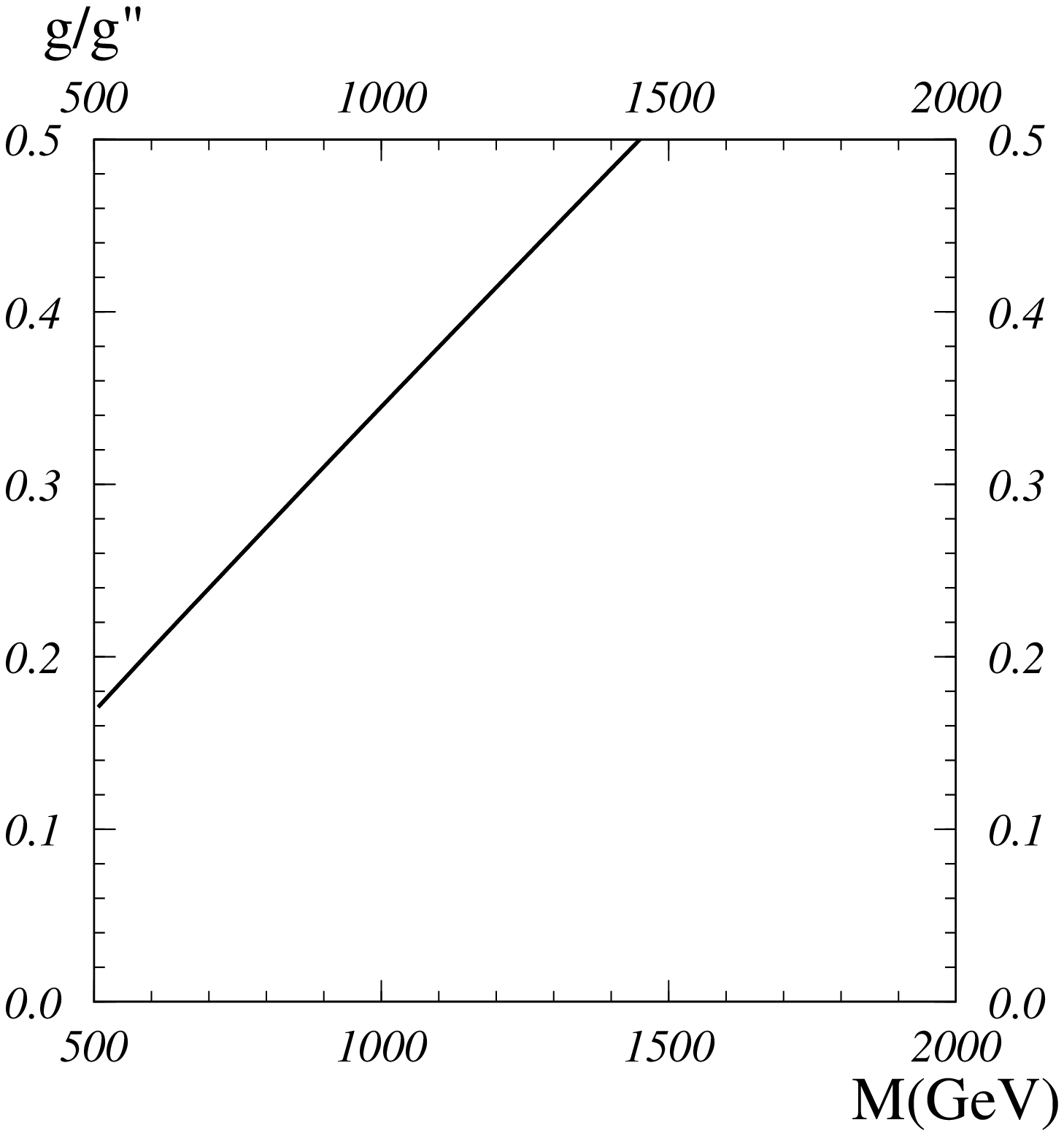}}
\smallskip
\noindent
{\bf Fig. 2} - {\it Degenerate  BESS model 90\% C.L. contour on the plane
($M$, $g/g''$) obtained by
comparing the theoretical predictions
to the experimental data. The allowed region is below the
curve.}
\label{degebo2}
\end{figure}

\section{Tests at $e^+e^-$ future colliders}

In presence of new vector resonances the annihilation channels $e^+e^-\to
f\bar f$, and $e^+e^-\to W^+W^-$ are much more efficient than the fusion
channels
because all the center of mass energy is used to produce the new particles.
For the BESS
model, since the new vector particles are strongly coupled, the channel
$W^+W^-$ is the dominant one \cite{yeit}. On the other hand, for the
degenerate BESS
model, due to its decoupling properties, the $W^+W^-$ channel gets depressed,
and the dominant one is $f\bar f$ \cite{dege}.

Our analysis, in the fermion channels, is
based on the following observables: the total hadronic
and muonic cross sections,
the forward-backward asymmetries in muons and $\bar b b$
 and, assuming a longitudinal polarization $P_e$ of the electron beam, the
left-right asymmetries in muons, $\bar b b$ and hadrons
\be
\sigma^{\mu},~~\sigma^h\nn
\ee
\be
A_{FB}^{e^+e^- \to \mu^+ \mu^-},~~ A_{FB}^{e^+e^- \to {\bar b} b}\nn
\ee
\be
A_{LR}^{e^+e^- \to \mu^+  \mu^-},~~A_{LR}^{e^+e^- \to {\bar b} b},~~
A_{LR}^{e^+e^- \to {had}}
\ee
We have assumed for $\sigma^h~(\sigma^\mu)$ a total error of $2\%~(1.3\%)$
\cite{dege}.
For the other observable quantities we assumed only statistical errors.

Concerning the $WW$ channel, we studied the following observables:
\be
{d\sigma \over {d\cos\theta}}(e^+ e^-\to W^+ W^-),~~~~~~
 A_{LR}^{{ e^+ e^- \to W^+ W^-}}
\ee
where $\theta$ is the center of mass scattering angle. We also
considered the possibility of measuring the final $W$ polarization
by using the
$W$ decay distributions, and we added
to our observables the longitudinal and transverse polarized $W$
differential cross sections and asymmetries.

\begin{figure}
\epsfysize=10truecm
\centerline{\epsffile{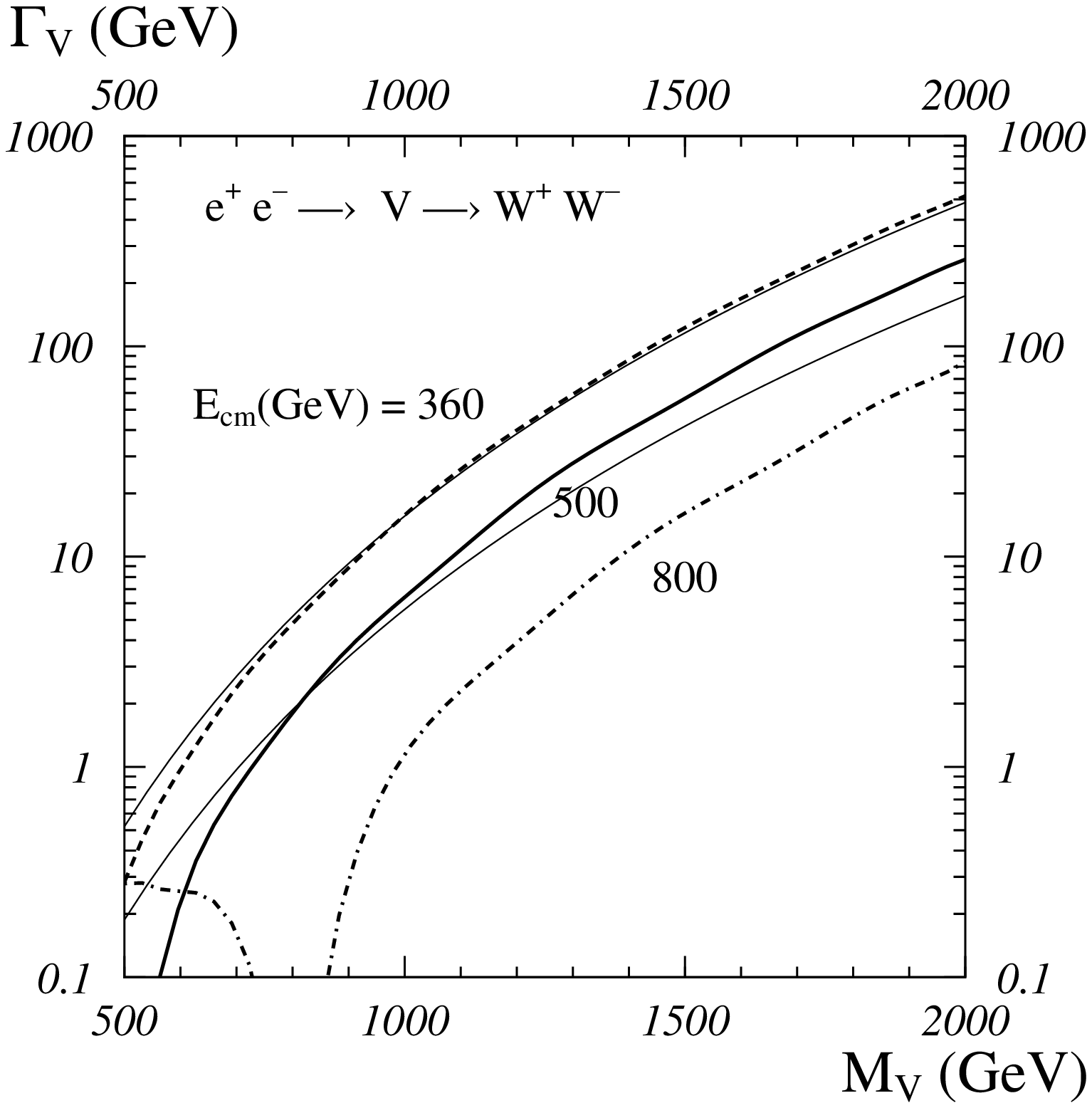}}
\noindent
\noindent
{\bf Fig. 3} - {\it BESS model $90\%$ C.L. contour in the $(M_V,\Gamma_V)$
plane
from measurements of differential cross sections and left-right
asymmetries; $W$ polarizations are reconstructed from the decay lepton
and quark jets. Energy and integrated luminosity are the following:
$\sqrt{s}= 360~GeV,~L=10~fb^{-1}$ (dashed)
 $\sqrt{s}= 500~GeV,~L=20~fb^{-1}$ (solid)
and  $\sqrt{s}= 800~GeV,~L=80~fb^{-1}$ (dash-dotted).
 The lower narrow solid line
is the limit from LEP measurements.
 The upper  narrow solid line is obtained by
considering the deviation with respect to the SM prediction and
assuming the LEP errors for the corresponding observables. The allowed region
is below the curves.}
\label{gamma}
\end{figure}

In order to get a clear signal for $W$ polarization reconstruction
we studied the channel for one $W$  decaying leptonically
and the other hadronically. We assumed
an effective branching ratio $B=0.1$ to take into account
the loss of luminosity from
beamstrahlung \cite{fuji}.

The analysis was performed by taking 19 bins in the angular region
restricted by
$|\cos\theta|< 0.95$ and assuming systematic errors
${{\delta B}/ B}=0.005$, $\delta {\cal L}/{\cal L}=1\%$ for the luminosity
and $1\%$ for the acceptance.

The machine options we considered are center of mass energies of
360, 500, 800 and 1000 $GeV$, and corresponding integrated luminosities of
10, 20, 50 and 80 fb$^{-1}$.

The results for the BESS model are shown in Fig. 3, where
the regions in the parameter space
$(M_V,\Gamma_V)$ which can be probed at
 $\sqrt{s}=360~GeV$, $L=10~fb^{-1}$ (dashed),
 $\sqrt{s}=500~GeV$, $L=20~fb^{-1}$ (continuous) and
 $\sqrt{s}=800~GeV$, $L=50~fb^{-1}$ (dot-dashed)
by measuring $W_{L,T} W_{L,T}$ differential cross
sections and left-right asymmetries are illustrated.
 The lower narrow solid line
is the limit from LEP measurements.
 The upper  narrow solid line is obtained by
considering the deviation with respect to the SM prediction and
assuming the LEP errors for the corresponding observables.

For instance for  $\sqrt{s}=500~GeV$, one is sensitive
for $M_V=2~TeV$ to $\Gamma_V\geq 250~GeV$,
for $M_V=1.5 ~TeV$ to $\Gamma_V\geq 60~GeV$, in agreement
with Barklow results \cite{bark}.

In conclusion measurements of  cross sections with
different final $W$ polarizations at a linear collider (LC) with
$\sqrt{s}=500~GeV$ improve LEP bound up to $M_V\sim 800~GeV$. At
$\sqrt{s}=800~GeV$ the sensitivity exceeds the LEPI bound for all
values of $M_V$.

\begin{figure}
\epsfysize=10truecm
\centerline{\epsffile{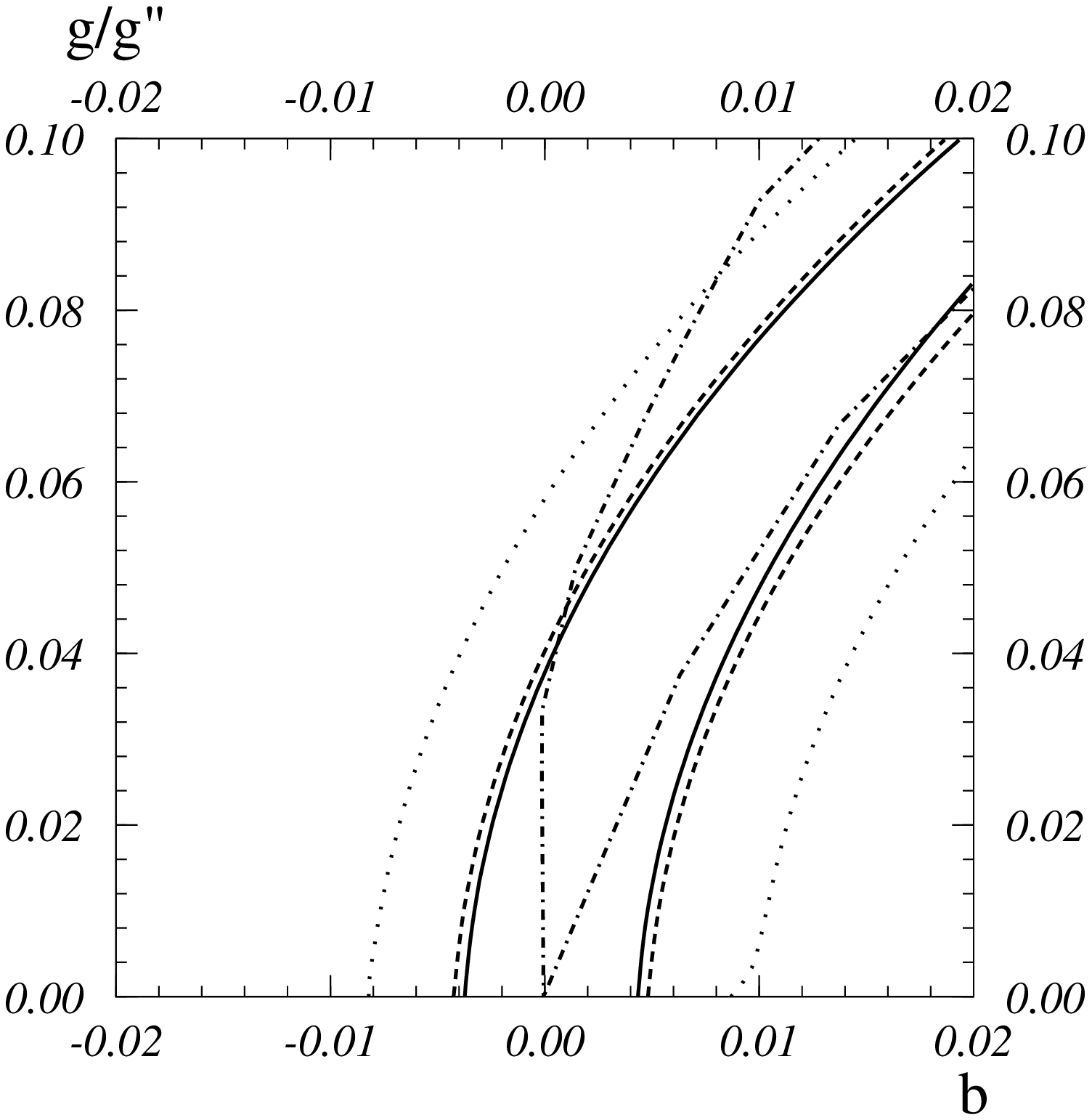}}
\noindent
\noindent
{\bf Fig. 4} - {\it BESS model 90\% C.L. contours in the
 plane $(b,g/\gs)$ for $M_V=1~TeV$.
The dotted  line corresponds to the bound from
the $WW$ differential cross section, the dashed  line from
 $W_{L}W_{L}$ differential cross sections and
the continuous line from  the differential
$W_{L,T}W_{L,T}$ cross sections and $WW$ left-right asymmetries at
a $500~GeV$ LC.
The dot-dashed line corresponds to the bound from the total cross
section
$pp\to W^\pm,V^\pm\to  W^\pm Z$ at LHC.
The allowed regions are  the internal ones.}
\label{cb1000}
\end{figure}

\begin{figure}
\epsfysize=10truecm
\centerline{\epsffile{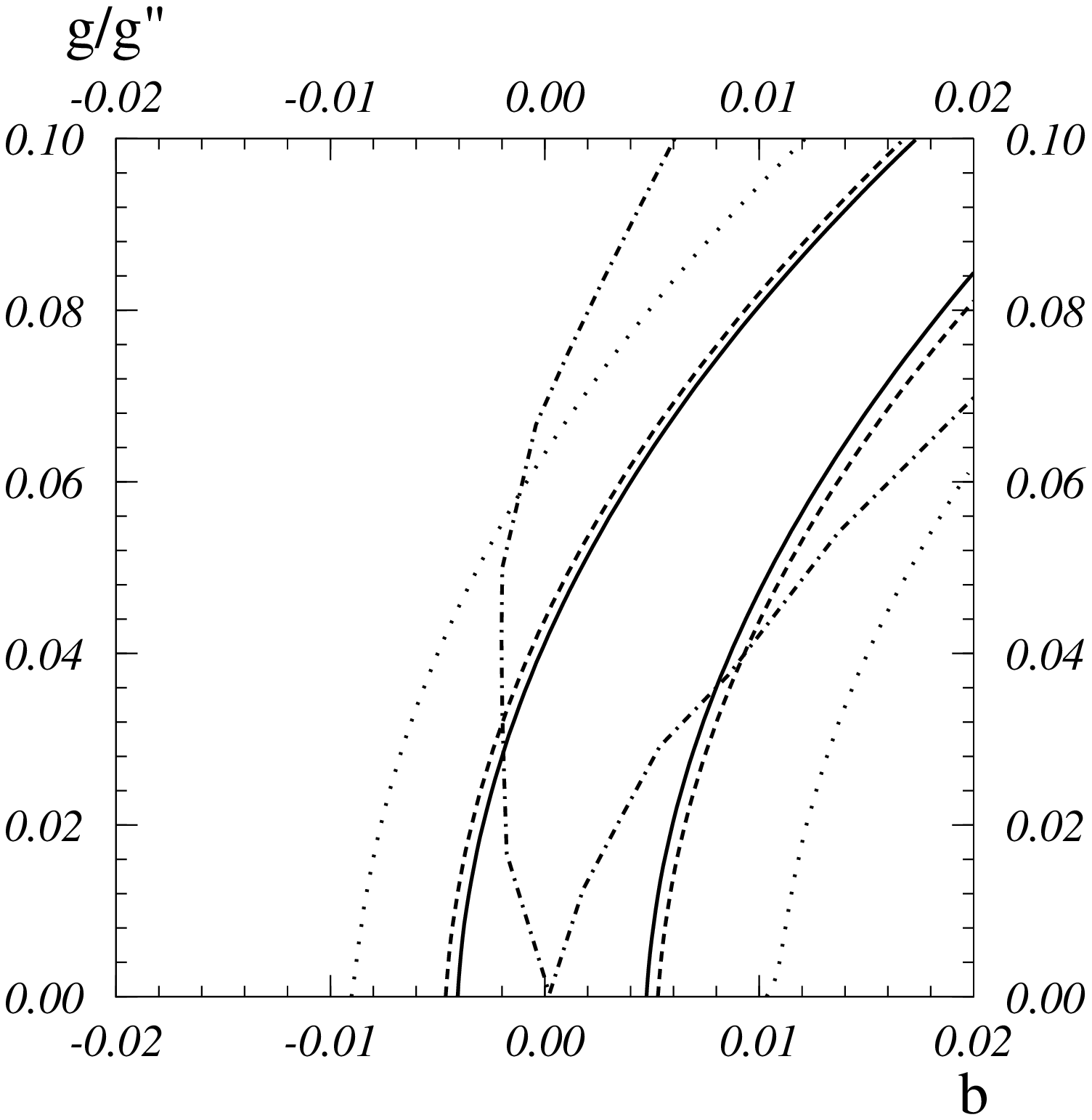}}
\noindent
\noindent
{\bf Fig. 5} - {\it BESS model 90\% C.L. contours in the
 plane $(b,g/\gs)$ for $M_V=1.5~TeV$.
The dotted  line corresponds to the bound from
the $WW$ differential cross section, the dashed  line from
 $W_{L}W_{L}$ differential cross sections and
the continuous line from  the differential
$W_{L,T}W_{L,T}$ cross sections and $WW$ left-right asymmetries at
a $500~GeV$ LC.
The dot-dashed line corresponds to the bound from the total cross
section
$pp\to W^\pm,V^\pm\to  W^\pm Z$ at LHC.
The allowed regions are  the internal ones.}
\label{cb1500}
\end{figure}

The comparison between
LHC and LC is shown in Fig. 4 (5) in the plane
$(b,g/\gs)$
for a LC of $\sqrt{s}=500~GeV$ and $L=20~fb^{-1}$ assuming
$M_V=1~TeV$  ($M_V=1.5~TeV$).
The dotted  line corresponds to the $90\%$ C.L.
bound from
the $WW$ differential cross section, the dashed  line to the bound
from $W_{L}W_{L}$ differential cross sections, and
the continuous line to the bound combining the differential
$W_{L,T}W_{L,T}$ cross sections and $WW$ left-right asymmetries.
The dot-dashed line represents the bound from LHC. The
allowed regions are the ones between the lines.
The bound from
LHC is obtained by considering the total cross section
$pp\rightarrow W^\pm,V^\pm\rightarrow  W^\pm Z\rightarrow \mu\nu \mu^+\mu^-$
and assuming that no deviation is
observed with respect to the SM within the experimental error.
Statistical and  systematic errors of $5\%$ have been
assumed.

This is for LHC the more efficient channel for the case of a vector
resonance strongly coupled to longitudinal $W$.
LHC can discover such vector resonances in a large region of
the parameter space up to masses $M_V=1.5-2~TeV$ in the channel
$pp\rightarrow W^\pm,V^\pm\rightarrow  W^\pm Z$
\cite{lhc1,CMS,ATLAS,bargerlhc}.

\begin{figure}
\epsfysize=10truecm
\centerline{\epsffile{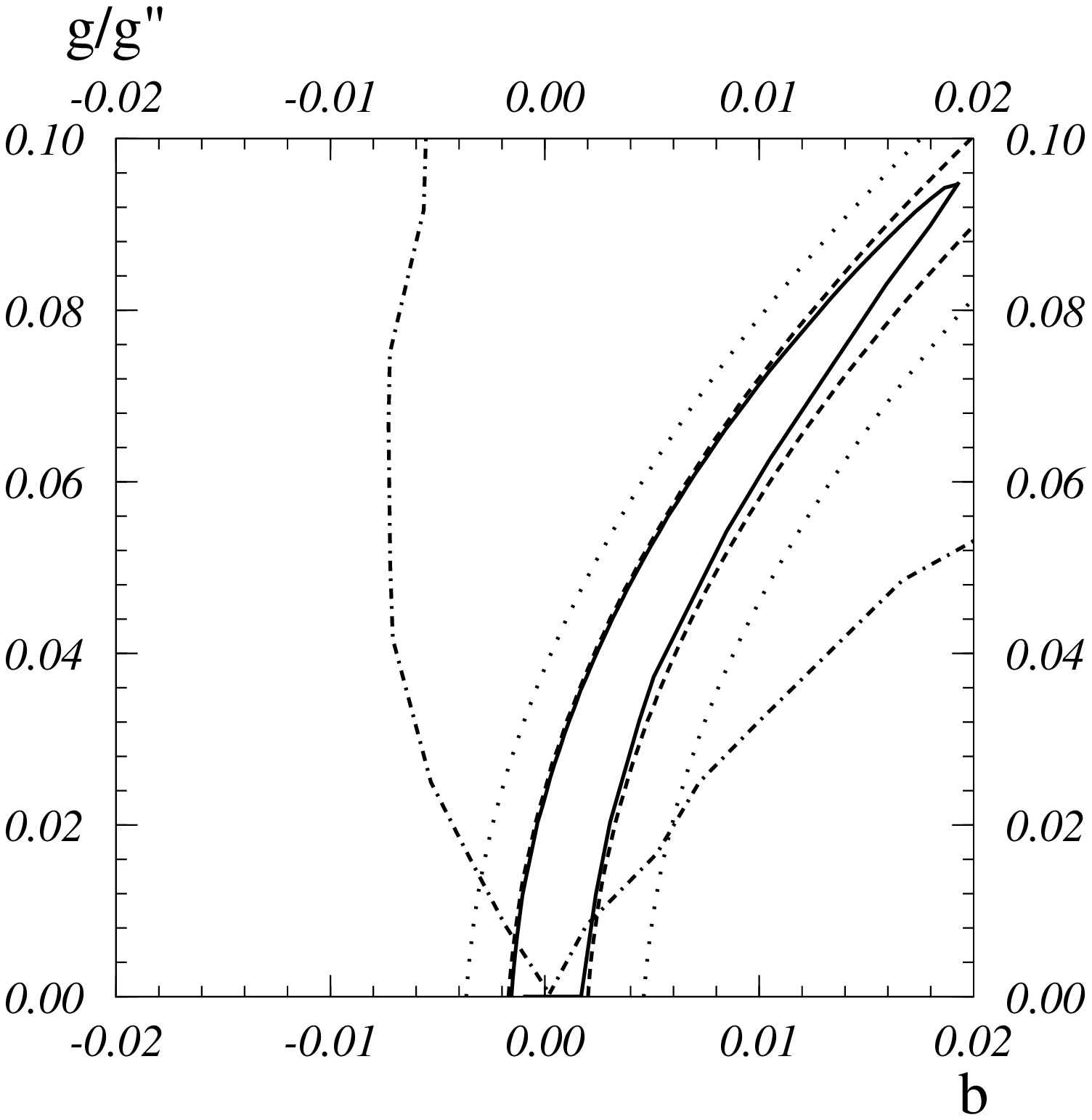}}
\noindent
\noindent
{\bf Fig. 6} - {\it BESS model 90\% C.L. contours in the
 plane $(b,g/\gs)$ for $M_V=2~TeV$.
The dotted  line corresponds to the bound from
the $WW$ differential cross section, the dashed  line from
 $W_{L}W_{L}$ differential cross sections and
the continuous line from  the differential
$W_{L,T}W_{L,T}$ cross sections and $WW$ left-right asymmetries at
a $800~GeV$ LC.
The dot-dashed line corresponds to the bound from the total cross
section
$pp\to W^\pm,V^\pm\to  W^\pm Z$ at LHC.
The allowed regions are  the internal ones.}
\label{cb2000}
\end{figure}

In Fig.6, $90\%$ C.L. contours in the
plane $(b,g/\gs)$ for $M_V=2~TeV$ are shown.
The dotted  line corresponds to the bound from
the $WW$ differential cross section, the dashed  line from
 $W_{L}W_{L}$ differential cross sections, and
the continuous line from  the differential
$W_{L,T}W_{L,T}$ cross sections and $WW$ left-right asymmetries at
a LC of $\sqrt{s}=800~GeV$ and $L=50~fb^{-1}$.
The dot-dashed line corresponds to the bound from the total cross
section
$pp\rightarrow W^\pm,V^\pm\rightarrow  W^\pm Z$ at LHC.

 The neutral channel
$pp\rightarrow \gamma ,Z,V\rightarrow  W^+ W^-$ suffers
of  background from $t\bar t$  production.
Nevertheless the new neutral vector bosons can be studied at LHC, by
considering their lepton decays, up to masses of the
order of $1~TeV$\cite{lhc2}.
Notice that while LHC is more
sensitive to the charged new vector bosons,
the LC is sensitive to the neutral ones (at least in the
annihilation channels).

As a general conclusion it seems that for models of new strong interacting
vector
resonances, if one is able to reconstruct the final $W$ polarization,
the measurement of polarized $e^+e^-\rightarrow W_LW_L$ gives rather strong
bounds
on the parameter space of the model.

Concerning the model with vector and axial-vector
resonances degenerate in mass, LHC is sensitive to
 the new particles in the channels
$pp\rightarrow W^\pm,L^\pm\rightarrow\mu\nu$ and
$pp\rightarrow\gamma,Z,L_3,R_3\rightarrow
\mu^+\mu^-$ up to masses of the order of $2~TeV$ as shown in
Fig. 7.

\begin{figure}
\epsfysize=10truecm
\centerline{\epsffile{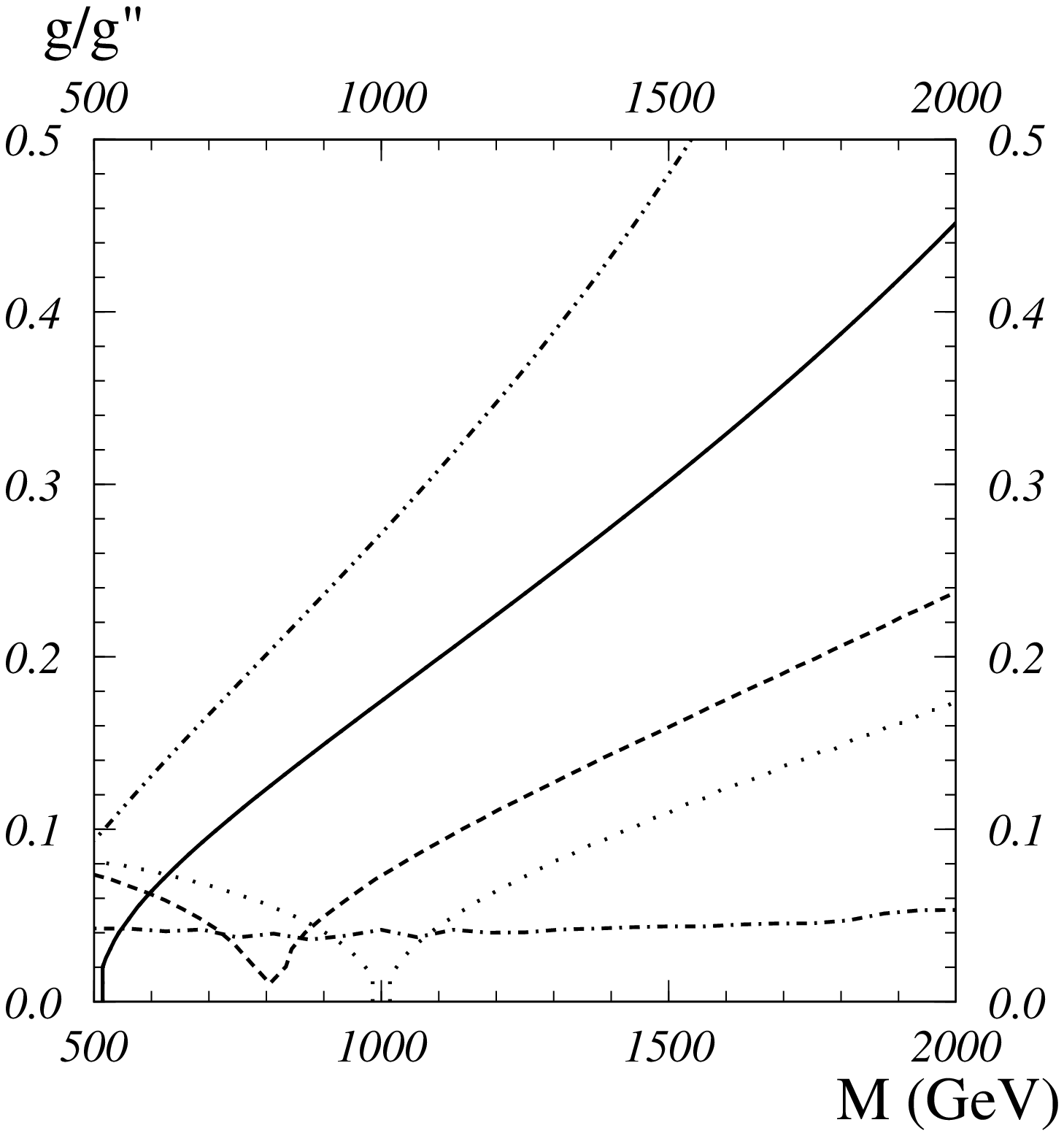}}
\noindent
{\bf Fig. 7} - {\it Degenerate BESS model
 90\%  C.L. contour on the plane ($M$, $g/g''$) from
$e^+e^-$ at different $\sqrt{s}$ values:  the dash-double-
dotted line represents
the limit from the
combined unpolarized observables at $\sqrt{s}=360~GeV$ with an
integrated luminosity of $L=10 fb^{-1}$,
the continuous line  at $\sqrt{s}=500~GeV$ and $L=20 fb^{-1}$,
the dashed  line  at $\sqrt{s}=800~GeV$ and $L=50 fb^{-1}$,
 the dotted line at $\sqrt{s}=1000~GeV$
and  $L=80 fb^{-1}$. The dot-dashed line represents
the limit from LHC. The allowed regions are below the curves.}
\label{deall}
\end{figure}

The bounds are obtained by assuming no
deviations with respect to the SM in the total
cross section $pp\rightarrow W^\pm,L^\pm\rightarrow\mu\nu$
within the experimental errors \cite{dege}.
Statistical and  systematic errors of $5\%$ have been
assumed. The bounds have to be compared with those coming
from LC's in  Fig.7.  The double dot-dashed
line represents the limit from the combined unpolarized observables at
$\sqrt{s}=360 ~GeV$ with an integrated luminosity of $L=10 fb^{-1}$;
the continuous line from  $\sqrt{s}=500 ~GeV$ and  $L=20 fb^{-1}$,
the dashed line from  $\sqrt{s}=800 ~GeV$ and  $L=50 fb^{-1}$
and the dotted line from
 $\sqrt{s}=1000$ GeV $L=80 fb^{-1}$. The dot-dashed line represents
the limit from LHC. Therefore, in the case of degenerate BESS, to be
competitive with
LHC, one
would need linear colliders of still higher energies than those considered.

\section{Conclusions}

We have studied tests for a possible strong electroweak sector by making
use of two models:
BESS, and degenerate BESS.

The importance of degenerate BESS as a model for a possible
strong electroweak sector is that it possesses a symmetry which allows it
to elude to
a high degree the present precision electroweak tests, notably from LEP. In
terms of the
spin one resonances present in the model this permits their allowed masses
to be lower
and their fermionic
couplings stronger than for ordinary BESS. Both models have charged as
well as neutral
resonances. For BESS, LC and LHC play complementary roles, in the sense
that while LHC
is suitable for the discovery of charged resonances, a linear collider is
important for
the neutral ones. Besides such complementarity, linear colliders in the
considered
ranges are found to be
quantitatively competitive to LHC if the strong electroweak sector is described
in terms of ordinary BESS.

The conclusions are different when degenerate BESS is
assumed to hold. This is due to the unique features of such a model. In
this case LHC
seems to be superior for testing the model when compared with the proposed
linear
colliders at the energies and luminosities considered. Due to more
favorable decay rate of
the resonances in fermions the conclusion holds not only for the charged
states but
also for the neutral states, differently than for ordinary BESS.

In view of the different possibilities that can be envisaged for a
possible strong electroweak sector, the general conclusion emerges that
both LHC and linear colliders will presumably play crucial roles in testing
for such a concept as a still viable alternative to other solutions of the
electroweak symmetry breaking problem, such as supersymmetry.
\newpage
\begin{center}
{\bf ACKNOWLEDGMENTS}
\end{center}
This work is part of the EEC project ``Tests
of electroweak symmetry breaking and future european colliders",
CHRXCT94/0579 (OFES 95.0200). A.D. acknowledges the support of a TMR
research fellowship of the European Commission under Contract nr.
ERB4001GT955869.

\end{document}